\documentclass[12pt]{article}
\usepackage{amsmath}
\usepackage{natbib}
\usepackage[dvips]{graphicx}
\usepackage[usenames,dvipsnames]{color}
\usepackage{url}

\begin{document}

\newcommand{\vecvar}{\vec}


\markright{Int. J. Remote Sens. 32(21): 6109-6132}

\title{Efficient statistical classification of satellite measurements}

\author{PETER MILLS$^{\ast}$\thanks{
$^\ast$Corresponding author: 
petey@peteysoft.org
1-613-833-3022
}\\
}

\maketitle

\begin{abstract}
Supervised statistical classification is a vital tool for satellite
image processing.  It is useful not only when a discrete result, such
as feature extraction or surface type, is required, but also for
continuum retrievals by dividing the quantity of interest into discrete
ranges.  Because of the high resolution of modern satellite instruments
and because of the requirement for real-time processing, any 
algorithm has to be fast to be useful.  Here we describe an
algorithm based on kernel estimation called Adaptive Gaussian Filtering
that incorporates several innovations
to produce superior efficiency as compared to three other popular methods: 
$k$-nearest-neighbour (KNN), Learning Vector Quantization (LVQ)
and Support Vector Machines (SVM).  This efficiency is gained with no
compromises:  accuracy is maintained, while estimates of the
conditional probabilities are returned.  These are useful not
only to gauge the accuracy of an estimate in the absence of its true
value, but also to re-calibrate a retrieved image and as a proxy for
a discretized continuum variable.  The algorithm
is demonstrated and compared with the other three
on a pair of synthetic test classes and to map the waterways of
the Netherlands.  Software may be found at: 
\url{http://libagf.sourceforge.net}.

\end{abstract}




\newcommand{\filtfunc}{f}

\section{Introduction}

Remote-sensing satellite instruments are returning increasing amounts of 
information about the Earth's surface and atmospheric state.  To be
useful, these data need to be processed to retrieve the quantities of
interest, ideally in real time or better.  Surface-detecting instruments,
such as Landsat, MODIS and AVHRR in the visible and infra-red and
AMSR-E and SSM/I in the microwave, are vital tools for mapping the globe, especially
when it is constantly shifting as in the case of sea ice \citep{Spreen_etal2008}.
Statistical classification can help determine what the instrument is
``seeing,'' 
whether that be crops, water, road or ice, 
underneath a given pixel or from a combination of pixels; for instance,
picking out crops of a particular type in a Landsat image \citep{Laue2004}.  
This type of
study can be useful both for producing maps and gathering statistics.
Even when a discrete type is not required, statistical classification
can still be useful for continuum retrieval and inversion by dividing
the quantity of interest into discrete ranges \citep{Mills2009}.

In supervised statistical classification, we are interested in determining
the class, $c$ of a test vector, $\vecvar{x}$, based on a series of known
input:output relations, $\lbrace \vecvar{x_i}:c_i \rbrace$, 
also known as training data.  
The vector, $\vecvar{x}$,
could correspond to a measurement vector, i.e. counts from
several channels of one or more satellite instruments, 
while the class, $c$ might correspond to a surface type, e.g. crops,
forest, field, road, water, etc.  The class is related to the inputs
via a conditional probability, $P(j|\vecvar{x})$, which is
discretely represented by the training data.  Normally, we seek the
the most likely class (maximum likelihood estimation):
\begin{equation}
c=\arg \underset {j} {\max} P(j | \vecvar{x})
\label{class_mle}
\end{equation}
thus we need some method of estimating the conditional probabilities.
This is the function of a kernel-density estimator 
\citep{Terrell_Scott1992}
and of a $k$-nearest-neighbours (KNN) method \citep{Michie_etal1994}.  
Other methods, such as Learning
Vector Quantisation (LVQ) \citep{Kohonen2000} 
and Support Vector Machines (SVM) \citep{kernel_intro} skip this step 
in favour of directly determining the class domains.
Note that kernel estimation should not be confused with methods based
on the ``kernel trick'' such as SVM, described in section \ref{sub_SVM}.

Because of the large amount of data involved and the necessity for real-time
processing, an important feature of modern satellite inversion algorithms,
including statistical classification, is speed.  
In \citet{Mills2009}, a method for statistical classification
called ``Adaptive Gaussian Filtering'' (AGF), based on kernel estimation,
 was briefly described and
applied to Advanced Microwave Sounding Unit (AMSU) data to discretely
retrieve water vapour in the upper troposphere.  Because this retrieval
required several days of AMSU swath data, comprising tens of millions of
individual measurements, the classification algorithm had to be fast.  
Here we describe the
algorithm more completely and demonstrate its superior performance 
compared to three other popular methods by  
first applying them to an artificial test case and then using them
to classify surface types in Landsat images.

The AGF algorithm incorporates several critical innovations
that make it extremely efficient without sacrificing
accuracy or the ability to estimate the conditional probabilities.
These are needed to set a definite confidence on the accuracy of an
estimate in absence of knowledge of its true value.
The following refinements are applied to a variable-bandwidth, kernel
density ``balloon'' estimator \citep{Terrell_Scott1992} based on Gaussian kernels.
First, the filter width is matched to the sample density using the
properties of the exponential function, avoiding unnecessary computation
of exponentials (section \ref{width_solve}).  Second, calculations are restricted to a set of 
$k$-nearest-neighbours found in $n \log k$ time with a binary tree (section \ref{binary_tree}).
Third, because probability estimates are continuous, 
a pre-trained model can be generated by searching for the class-
borders with guaranteed, super-linear convergence (sections \ref{class_borders}
and \ref{rootfinding}).  Fourth, using the
pre-trained model, the conditional probabilities are interpolated from
gradients at the class-border (section \ref{cond_prob_int}).

\section{Essential description of the algorithm}

\subsection{Adaptive Gaussian filtering}
\label{AGF_intro}

The $k$-nearest-neighbours (KNN) is a well-known and effective technique of
estimating probability densities and performing classifications.
It works by picking from the training data the $k$ samples nearest the test point 
and determining the class by voting \citep{Michie_etal1994}.
A simple refinement to this method would be to weight the samples
according to distance as in a simple linear filter.
Given a set of points, $\lbrace \vecvar{x_i}\rbrace$, 
the probability density function (PDF) of a test point, $\vecvar x$, 
may be estimated as follows:
\begin{eqnarray}
P(\vecvar x) & \approx & \frac{W}{n N} \label{pdf_est} \\
W & = & \sum_{i=0}^n w_i
\label{W_def}
\end{eqnarray}
where $w_i$ is the weight of the $i$th sample, $n$ is the number
of samples and $N$ is a normalisation coefficient.  Its justification
is a Monte Carlo integration with importance sampling \citep{nr_inc2}, 
except that here we solve for the importance distribution by multiplying it
with a peaked, but otherwise essentially arbitrary function, and
assume that it is roughly constant.  

The magnitude
of the weights must decrease with distance and in the isotropic
case they are given by:
\begin{eqnarray}
w_i & = & \filtfunc(d_i) \\
d_i & = & | \vecvar x - \vecvar{x_i} |
\end{eqnarray}
where $\filtfunc$ is a filter function and $d_i$ is the 
distance of the $i$th
sample from the test point.  The upright brackets denote a metric,
typically Cartesian, although in theory any metric could be used.
In practise, it is often simplest and most efficient to first re-
scale or otherwise transform the variables so that a Cartesian metric
is then appropriate.  The normalisation coefficient will be given by:
\begin{equation}
N =  \int_A \filtfunc(|\vecvar x - \vecvar y|) \mathrm{d} \vecvar y
\label{norm_def}
\end{equation}
where $A$ is the domain of the samples.
This technique is known as a fixed-bandwidth kernel-density estimator \citep{Terrell_Scott1992}.

A natural choice for $\filtfunc$ would be a Gaussian:
\begin{equation}
\filtfunc(r) = \exp \left (- \frac{r^2}{2 \sigma^2} \right )
\label{Gaussian_filter}
\end{equation}
where $\sigma$ is the filter width and $N = (2 \pi)^{D/2} \sigma^D$
with $D$ as the number of dimensions.  Using a fixed filter width
may mean that in regions of low density, all samples will fall in
the tails of the filter with very low weighting, while regions of
high density will find an excessive number of samples in the central
region with weighting close to unity.  Thus we vary the
filter width according to density so that an optimal number of
samples falls within the central region.

According to the definition of the probability density, the local
average point spacing will be given as follows:
\begin{equation}
\delta = \frac{1}{\left [ n P(\vecvar x) \right ]^{1/D}}
\end{equation}
We wish to vary the filter width so that it matches the point spacing:
\begin{equation}
\sigma^{(opt)} = k_1 \delta = \frac{k_1}{\left [ n P(\vecvar x) \right ]^{1/D}}
\end{equation}
where $k_1$ is a coefficient.
Substituting the approximated PDF from (\ref{pdf_est}) through
(\ref{Gaussian_filter}) produces the
following that must be solved for $\sigma^{(opt)}$:
\begin{equation}
W (\sigma^{(opt)}) \approx k_1^D (2 \pi)^{D/2} = \mathrm{const.}
\label{opt_filt2}
\end{equation}
Thus, correctly selecting a fixed value for $W$ (call it $W_c$) in
(\ref{W_def}) should produce an ``optimal'' (as we have defined it)
filter width.  
The proof is valid for all functions for which $N \propto \sigma^D$
which is true for all functions of the form $f(r/\sigma)$ for
which the integral in (\ref{norm_def}) exists.

The quantity $W_c$ may be thought of as roughly equivalent to $k$ in a
KNN scheme.  Here we have created what is known as
a variable-bandwidth kernel-density ``balloon'' estimator:
variable-bandwidth means that the filter width
is varied depending on the location in the sample space, 
while ``balloon'' means that it is varied based on the location
of the test point, not the training samples (as in a ``pointwise estimator''),
thus for a given estimate the same filter width is applied 
to all training samples~\citep{Terrell_Scott1992}.

To use the method for classification, we first estimate the
conditional probability as follows:
\begin{equation}
P(m | \vecvar x) \approx \frac{1}{W} \sum_{i, c_i=m} w_i
\label{cond_prob_est}
\end{equation}
where $c_i$ is the class associated with the $i$th sample.

\subsection{Finding the class borders}

\label{class_borders}

The chief advantage of this scheme over a KNN is that
it produces results that are both continuous and differentiable.  
Both properties are desirable in that they allow us to search for a unique
border between the classes (discrimination border) and hence make more rapid classifications. 
Assuming that there are
only two classes, the difference in their conditional probabilities is:
\begin{equation}
R(\vecvar x) = P(2 | \vecvar x) - P(1 | \vecvar x) \approx \frac{1}{W}
\sum_i (2 c_i - 3) w_i
\label{Rdef}
\end{equation}
where $1$ and $2$ are the classes and $c_i$ is the class of the $i$th sample.
The borders are found by setting this expression to zero,
$R(\vecvar x)=0$.  With an adaptive Gaussian filter, the derivative becomes:
\footnote{Superscripts have been omitted.  For a derivation of this equation (which does not appear
in the original paper) please refer to Appendix \ref{appendix}.}
\begin{equation}
\frac{\partial R}{\partial x_j} \approx \frac{1}{\sigma^2 W_c}
	\sum_i w_i (2 c_i - 3) \left [x_{ij}-x_j - d_i^2 \frac{\sum_k
	w_k (x_{kj}-x_k)} {\sum_k d_k^2 w_k} \right ] 
\label{class_grad}
\end{equation}
where $x_{ij}$ is the $j$th coordinate of the $i$th sample, while $x_j$ is the
$j$th coordinate of the test point.  
Derivatives will be necessary both 
for estimating the class of the test point (see below) 
and then extrapolating the conditional probabilities (section \ref{cond_prob_int})
as well as useful (though not essential) 
for searching out the discrimination border (section \ref{rootfinding}).

The border may be sampled as many times as necessary,
giving a set of vectors, $\lbrace \vecvar{b_i} \rbrace$, 
along with their corresponding gradients,
$\lbrace \nabla_{\vecvar x} R |_{\vecvar x=\vecvar{b_i}} \rbrace$.  The class of a test point,
$\vecvar x$, is calculated as follows:
\begin{eqnarray}
j & = & \arg \underset{i}{\min} | \vecvar{b_i} - \vecvar x | \label{jeq}\\
p & = & (\vecvar x - \vecvar{b_j}) \cdot \nabla_{\vecvar x} R |_{\vecvar{x}=\vecvar{b_j}} \label{peq} \\
c & = & ( 3 + p/|p| ) / 2 \label{ceq}
\end{eqnarray}
where $c$ is the class.  The specific procedure used to sample the
class border will be described in section \ref{rootfinding}.

Note that it is easy to generalise a two-class classification to multiple
classes, although the best method of doing so will be highly problem-dependent.

\subsection{Extrapolating the conditional probabilities}

\label{cond_prob_int}

The value of $R$ may be extrapolated to the test point.
Consider a pair of one-dimensional classes composed of two equal-sized 
Gaussians of width $s$ separated by a distance $2a$ 
with the class border lying at $b$.
Let $\tilde R$ be the difference between the conditional probabilities:
\begin{eqnarray}
  \tilde R(x) & = & P(2|x) - P(1|x) \\
  & = & \frac{P(2, x) - P(1, x)}{P(1, x) + P(2, x)} \\
  & = & \frac{e^{-\frac{(x-b+a)^2}{2 s^2}} -
                e^{-\frac{(x-b-a)^2}{2 s^2}}}
                {e^{-\frac{(x-b+a)^2}{2 s^2}} +
                e^{-\frac{(x-b-a)^2}{2 s^2}}}
\end{eqnarray}
Simplifying:
\begin{equation}
\tilde R (x) = \tanh \left (\frac {a(x-b)}{s^2} \right )
\label{cext:r1sim}
\end{equation}
To approximate the difference in
conditional probabilities, $R(x)$, between an arbitrary pair of (well-behaved)
1-D classes, we fit $\tilde R(x)$ to $R(x)$ by setting:
\begin{eqnarray}
\tilde R(b) = R(b) & = & 0 \\
\left . \frac{\mathrm{d}\tilde R}{\mathrm{d}x}\right |_{x=b} = \frac{a}{s^2} & = & \left . \frac{\mathrm{d}R}{\mathrm{d}x} \right |_{x=b}
\end{eqnarray}
We can approximate $R$ for a pair of multi-dimensional classes 
in the same way:
\begin{equation}
R(\vecvar x) \approx \tilde R(\vecvar x) \equiv \tanh p
\label{confidence_est}
\end{equation}
It is easy to show:
\begin{eqnarray}
\tilde R(\vecvar{b_j}) & = & 0 \\
\nabla_{\vecvar x} \tilde R |_{\vecvar{x}=\vecvar{b_j}} & = &\nabla_{\vecvar x} R|_{\vecvar{x}=\vecvar{b_j}}
\end{eqnarray}
The approximation assumes that the PDFs of the classes are roughly Gaussian near
the discrimination border.

\section{Refinements}

\subsection{Solving for the filter width}

\label{width_solve}

The properties of the exponential function can be used to 
solve for the filter width by iteratively squaring the weights.  Let the $j$th
weighting coefficient of the $i$th sample be given as:
\begin{equation}
w_i^{(j)} = \exp \left ( \frac{d_i^2}{2 ( \sigma^{(j)})^2} \right )
\end{equation}
where $\sigma^{(j)}$ is the $j$th filter width.  Each subsequent iterate is
defined as the square of its previous:
\begin{equation}
w_i^{(j)} = \left (w_i^{(j-1)} \right )^2
\end{equation}
consequently the $j$th filter width, $\sigma^{(j)}$, obeys the following
recursion relation:
\begin{equation}
\sigma^{(j)} = \frac{\sigma^{(j-1)}}{\sqrt{2}}
\end{equation}
Following our super-scripting convention, $W^{(j)}$ is defined as:
$W^{(j)} = \sum_i w_i^{(j)}$ with the final iteration of $j$, call it
$f$, defined such that:
\begin{equation}
W^{(f)} \le W_c
\end{equation}
Obviously, the initial filter width, $\sigma^{(0)}$, must be chosen to be larger
than the optimal, for instance by taking the total variance of the data.

The final filter width is approximated by exponential interpolation to
the target total weight:
\begin{eqnarray}
k_2 & = & \frac{\log W_c + \log W^{(f)} - 2 \log W^{(f-1)}}{2\left
	(\log W^{(f)} - \log W^{(f-1)}\right )} \\
\sigma^{(opt)} & \approx & \frac{\sigma^{(f)}}{\sqrt{k_2}}
\end{eqnarray}
The advantage of this scheme is that the exponentials, the most expensive
part of the calculation, are computed only twice for each
training sample.  This is in contrast to a root-finding algorithm, which
would need three computations at minimum.

\subsection{Restricting calculations to $k$-nearest-neighbours}
\label{binary_tree}

\begin{figure}
\includegraphics[angle=90,width=0.95\textwidth]{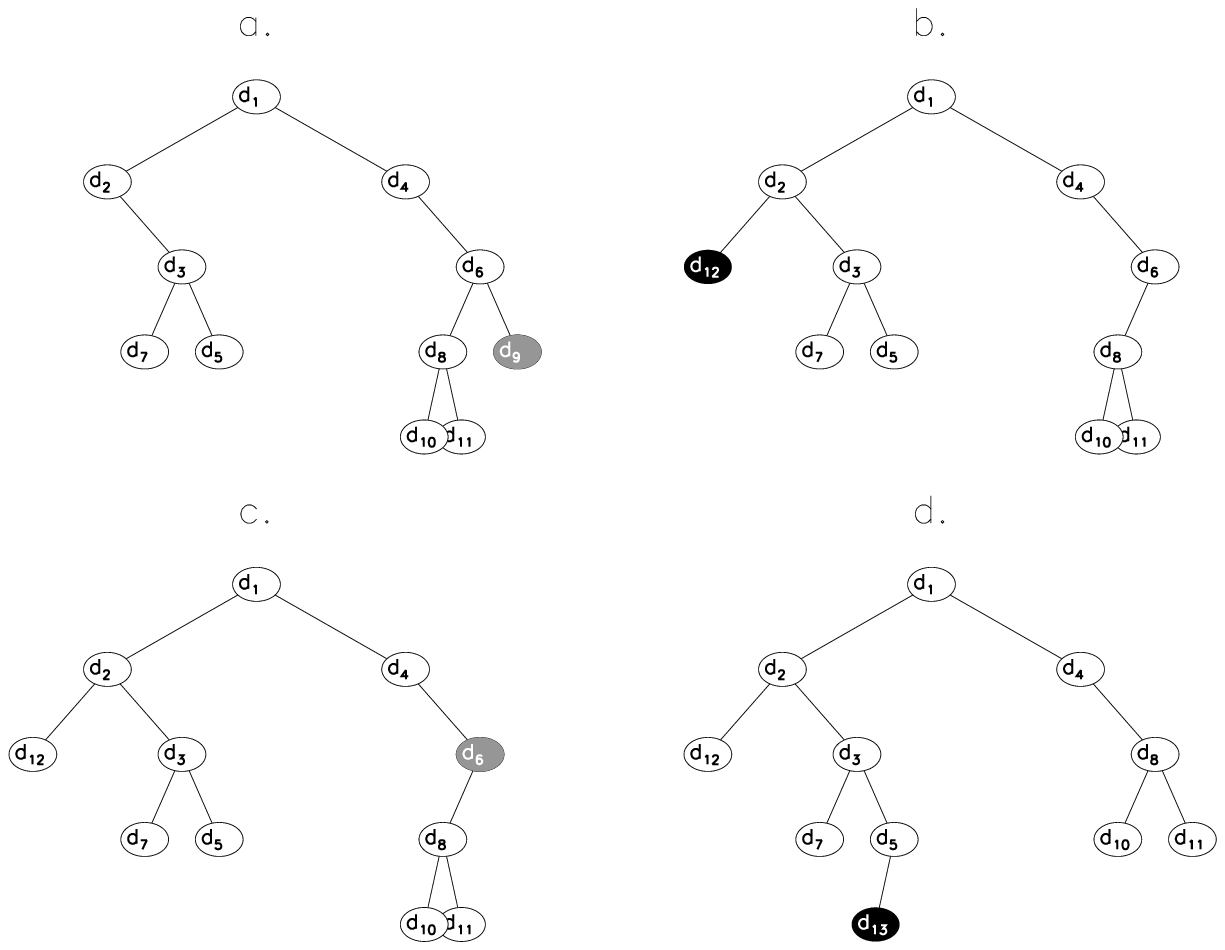}
\caption{Demonstrating the use of a binary tree to select the 10 least
elements from a list of distances, $\lbrace d_i \rbrace$.  
In each illustration, elements to the left are
smaller, while elements to the right are larger.
In (\textit{a}), the tree has been filled with the first eleven members from the
list and the largest element in the tree is marked for removal.
In (\textit{b}), the operation is completed and the 12th member from the list is added.
In (\textit{c}), the largest element in the tree is once again marked for removal.
In (\textit{d}), the operation is completed and the 13th member from the list is added.}
\label{kgreatest}
\end{figure} 

Although the filter function, $\filtfunc$, is applied in theory to all the samples,
in practice only a subset of nearest neighbours will make a significant contribution
to the final result.  We can find $k$ nearest neighbours in $n \log k$ time using a
binary tree.  The procedure is illustrated in figure \ref{kgreatest} and described
in words in the following paragraph. 

All the distance must be calculated and the
$k$ least of these will correspond to the desired neighbours. 
The first $k+1$ elements are arranged in a binary tree.  The largest element will be
the rightmost in the tree and must be deleted.
This can be done in roughly $\log k$ time by
traversing the tree from its root.
A new element is then added to the tree, also in $\log k$ time, and the largest once again deleted.   
The procedure is repeated until all the elments
in the list have been added to the tree and the only ones remaining are
the $k$ least \citep{Knuth1998}.

While it might appear that repeatedly deleting the largest element will
produce an unbalanced tree, actual tests suggest that this is not the case.
There are at least two mitigating factors.
First, the greatest element is not always a ``leaf'' or lowermost element, but
is itself often a node, thus an entire sub-tree will take its place
as figures \ref{kgreatest}(\textit{c}) and (\textit{d}) exemplify.  Second, new elements are constantly being
added to the tree.  These will tend to be larger on average than those already occupying
it since we are selecting for the $k$ least.

\subsection{Root-finding}
\label{rootfinding}

\begin{figure}
\includegraphics[width=0.95\textwidth]{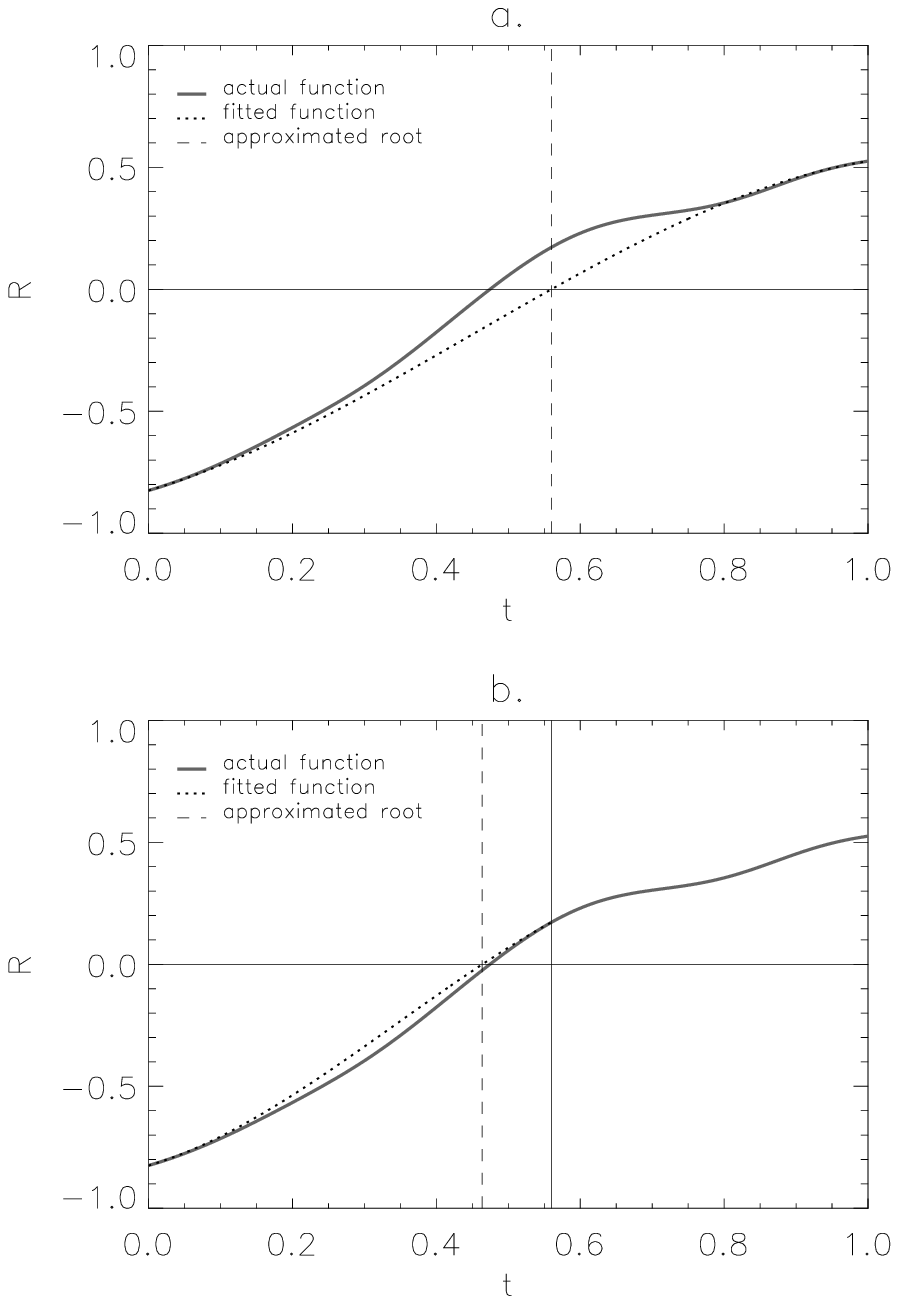}
\caption{Demonstration of the root-finding algorithm.  In a., 
the root is bracketed for the first time.  A third-order polynomial
is fitted and the root approximated by solving the cubic equation.
In b., the function has been re-bracketed with the new root
and a new polynomial fitted between the new brackets.}
\label{root_demo}
\end{figure}

To sample the class border, the following procedure was employed:  
pick two points at random,
$\vecvar{x_1}$ and $\vecvar{x_2}$, belonging to classes 1 and 2 respectively.  We
define $\vecvar v$ as a direction vector between the two points, e.g.:
\begin{equation}
\vecvar v = \vecvar{x_2} - \vecvar{x_1}
\end{equation}
and solve the following for t:
\begin{equation}
R(\vecvar{x_1} + \vecvar v t) = 0
\end{equation}
to find a point,  $\vecvar b = \vecvar{x_1} + \vecvar v t|_{R=0}$, 
randomly located on the class border.

This reduces the $D$-dimensional root-finding problem to only one dimension,
with the root already bracketed, thus it can be found with considerable
certainty.
The root of a one-dimensional function is considered bracketed when we
have two values of the independent variable for which the function evaluates
to opposite signs, therefore at least one root must lie between \citep{nr_inc2}. 
The derivatives, $\mathrm{d} R/\mathrm{d} t = \nabla_{\vecvar x} R \cdot \vecvar v$, are used
as an aid to increase the speed of convergence.
If both the value and the first derivative of the function 
are known at two locations, then it is possible to fit a unique, third-order
polynomial.  The root is estimated by equating the polynomial to zero,
and the true root re-bracketed with this new estimate
-- see figure \ref{root_demo}.
The procedure is repeated until the true root is found to within
a certain tolerance.  This combines the fast convergence of a Newton's method
with the numerical stability of a bisection algorithm \citep{nr_inc2},
hence we term the method, ``supernewton.''
The GNU scientific library (GSL) is used both to 
solve the cubic and to fit the function by
solving the rank four linear system using Householder
transformations \citep{gsl_ref}.

\section{Comparison with competing algorithms using an artificial dataset}

\subsection{A pair of test classes}
\label{test_classes}

\begin{figure}
\begin{center}
\resizebox{1\textwidth}{!}{
  \includegraphics{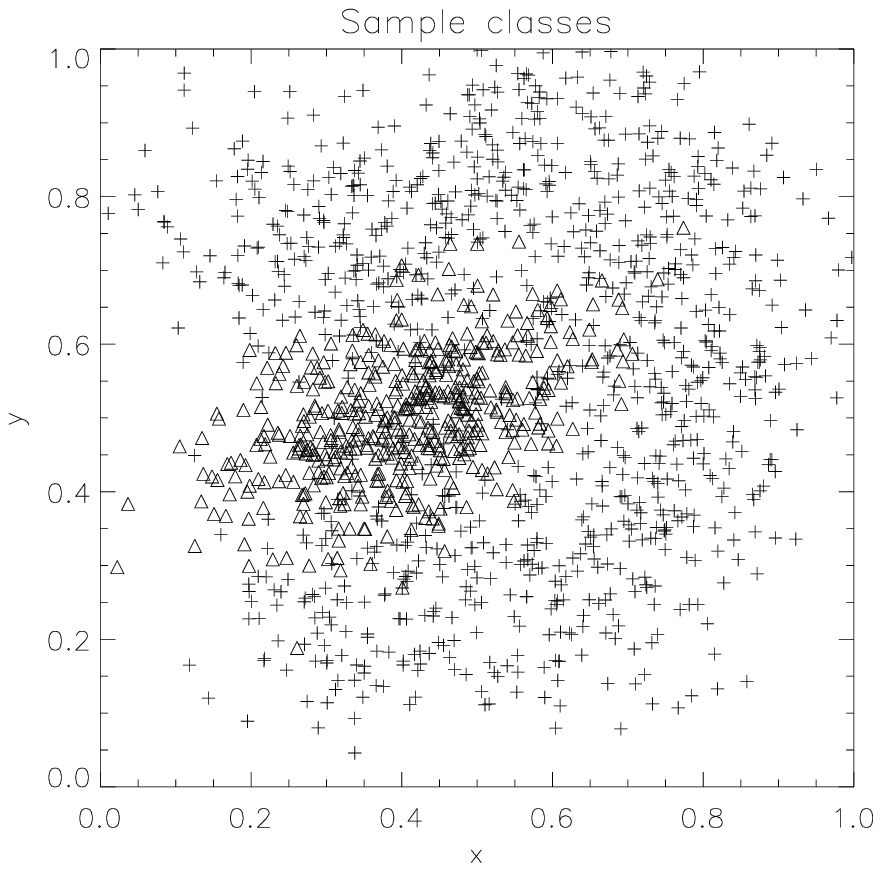}}
  \caption{A pair of synthetic test classes}
  \label{cls_exmpl}
\end{center}
\end{figure}

\begin{table}
\caption{Table of values used to define the ``spine'' of the distribution defining
the second of the two synthetic test classes.}
\label{spine}
\begin{center}
\begin{tabular}{|cc|}
\hline
x & y \\
\hline \hline
0.17 & 0.79 \\
0.36 & 0.70 \\
0.51 & 0.84 \\
0.70 & 0.86 \\
0.76 & 0.68 \\
0.77 & 0.48 \\
0.68 & 0.32 \\
0.46 & 0.28 \\
0.24 & 0.26 \\
\hline
\end{tabular}
\end{center}
\end{table}

The AGF method was tested on a synthetic dataset consisting of the pair of
classes shown in figure \ref{cls_exmpl}.  The first class, illustrated using
triangles, is a simple, two-dimensional normal distribution:
\begin{eqnarray} 
P(\vecvar x | 1) & = & \frac{1}{2 \pi \sigma_1 \sigma_2} 
	\exp \left \lbrace - \frac{1}{2}\left 
	[\left (\frac{x^\prime}{\sigma_1}\right )^2 + 
	\left (\frac{y^\prime}{\sigma_2} \right )^2 \right ] \right \rbrace
\label{pdf_sc1a} \\
\vecvar x^\prime & = & \left [\begin{array}{cc} 
	\cos \theta & \sin \theta \\
	-\sin \theta & \cos \theta
\end{array} \right ] \cdot (\vecvar x - \vecvar{x_0}) 
\label{pdf_sc1b} 
\end{eqnarray}
where $\vecvar{x_0}$ is the centre of the distribution, $\sigma_1$ and $\sigma_2$ are
the widths of its major and minor axes respectively, $\theta$ is the 
angle of the major axis and $\vecvar x^\prime = (x^\prime, y^\prime)$.

To produce a non-trivial interface between the two classes, 
the second one has a more intricate design.
A set of points defining the ``spine'' of the distribution was chosen
-- see Table \ref{spine}.
Individual samples are first interpolated a random distance along 
the curve so defined using a cubic spline\citep{nr_inc2,gsl_ref}.
By displacing this point a random distance in a random direction, 
the final location of the sample is determined.

Analytic or semi-analytic 
(e.g. using numerical quadradure to perform the integration)
values for the probability densities of
the second class may be calculated as follows:
\begin{equation}
P(\vecvar x | 2) = \frac{1}{s_{max}}\int_0^{s_{max}} Q(|\vecvar x - \vecvar g(s)|) \mathrm{d} s
\label{pdf_sc2}
\end{equation}
where $\vecvar g$ is the line segment
defining the spine, $s$ is the path along it, $s_{max}$ is its
length and $Q(r)$ is a circularly symmetric
PDF governing the offset distance from the backbone.
For the dataset shown in figure \ref{cls_exmpl}, 
a Gaussian (of width 0.1) was once 
again employed for $Q$.

\subsection{Competing algorithms}

To test the effectiveness of the AGF classification algorithm,
it was compared with three other, popular methods.
These are: $k$-nearest-neighbours (KNN), Learning Vector Quantisation (LVQ)
and Support Vector Machines (SVM) and are briefly described
below.

\subsubsection{$K$-nearest-neighbours}

The KNN is one of the simplest and most robust
classification methods available.  It consists of finding
the $k$ training samples that are closest to the test point and then
counting how many there are of each class.  The class with the most
samples is the class of the test point.

Our method reduces to a KNN by using a filter function that
is one out to the filter width and zero everywhere else:
\begin{equation}
\filtfunc(r) = \left \lbrace \begin{array}{lr}
\filtfunc(r) = 1 , & r < \sigma \\
\filtfunc(r) = 0 , & r \ge \sigma
\end{array} \right .
\end{equation}
where $\sigma$ is the filter width.

\subsubsection{Learning Vector Quantisation}

LVQ works by building up a set of ``codebook vectors''
whose density matches the difference between the densities of the
two classes.  The vectors will be labelled based on which class they
fall within.  The class of a test point will be given by the class
label of the nearest codebook vector.

The training process is performed as follows:  a set of codebook
vectors are initialised at random points and assigned class labels.  
The codebook vectors are randomly compared with the training samples.
If the labels of the two match, then the codebook vector is moved
closer to the training sample.  If they don't, then it is moved further
away\citep{Kohonen2000,LVQ_PAK}.

\subsubsection{Support Vector Machines}

\label{sub_SVM}

In SVM, a single hyperplane is drawn that best
divides the two classes.  Classifications are done as in equations
(\ref{jeq})--(\ref{ceq}) based on a dot product with the normal
to the class border.  The border is fitted by minimizing the classification
error.  Obviously, using a single straight line to divide the two
synthetic sample classes will produce poor results. 
Results may be improved by 
adding variables derived from
the originals thus expanding the dimension of the space, 
much as one fits a nonlinear function, e.g. a polynomial, 
by performing a linear fit on the set of variables formed 
by transforming the independent variables with a set of basis functions.

The so-called kernel trick can be applied to most algorithms
that use scalar products and is based on the observation that certain
mathematical operations applied to a scalar product will expand the
variables in a set of basis functions, 
without having to explicitly calculate them.
For example, consider the squared dot product of two two-dimensional 
variables \citep{kernel_intro}:
\begin{eqnarray}
(\vecvar x \cdot \vecvar y)^2 & = & [(x_1, x_2) \cdot (y_1, y_2)]^2 \\
& = & x_1^2 x_2^2 + 2 x_1 y_1 x_2 y_2 + x_2^2 y_2^2 \\
& = & (x_1^2, \sqrt{2} x_1 x_2, x_2^2) \cdot (y_1^2, \sqrt{2} y_1 y_2, y_2^2)
\end{eqnarray}
    
\subsection{Comparison results}

\begin{table}
  \begin{center}
  \caption{Comparison of parameters used in the KNN, LVQ, AGF
	and AGF borders classification algorithms}
  \begin{tabular}[h]{|l|l|l|l|}\hline
    KNN & LVQ & AGF & AGF borders \\ \hline \hline
    $k = 101$ & $\alpha_0=0.1$ & $W_c = 100$ & $W_c=100$ \\
      & $n_t=75000$ & $k=1000$ & $k=1000$ \\
      & $n=1000$ & & $n=250$ \\ 
      & & & $\epsilon=0.0001$ \\ \hline
  \end{tabular}
  \label{parm_table}
  \end{center}
\end{table}

\begin{table}
  \begin{center}
  \caption{Parameters used in the SVM classification algorithm}
  \begin{tabular}[h]{|ll|}\hline
    implementation: & LIBSVM \\
    type: & C-SVC \\
    kernel basis function: & $\exp(\gamma |\vecvar{x_i} - \vecvar{x_j}|^2)$ \\
    $\gamma$: & $0.5$ \\
    $C$ (cost): & $100$ \\ 
    $\epsilon$: & $0.001$ \\ \hline
  \end{tabular}
  \label{SVM_parm}
  \end{center}
\end{table}

Random synthetic datasets composed of 5000 samples of the first class
and 10000 samples of the second class were created as needed
using the algorithms described in section \ref{test_classes}.
Test datasets composed of three-thousand (3000) members
were created separately and had no fixed ratio
between the number of members in each class.  Rather the
ratio of the training data was used to randomly select which
class was sampled.

The algorithm was also compared with an analytical classification
scheme that compares the results of equations (\ref{pdf_sc1a}) and 
(\ref{pdf_sc2}) applied to the test point.  This has the advantage
of quantifying the limit in accuracy of any classification algorithm,
as well as returning the conditional probabilities.

Parameters for the KNN, LVQ, AGF and AGF borders techniques are
compared in Table \ref{parm_table}.  For the LVQ method, Kohonen's
so-called optimised LVQ 1 (OLVQ1) method was used.  $n_t$ is the 
number of training cycles, while $n$ is the number of codebook
vectors for the LVQ method and the number of border samples for
the AGF border classification method.  
Note that AGF can be applied directly 
without searching for the class borders
using equation (\ref{cond_prob_est}).
Also, performing the AGF classifications with all the data would
be too slow, so the $k$-nearest-neighbours supplying the most 
weight are selected before
applying the algorithm.  This is done in $n \log k$ time using a
binary tree as described in section \ref{binary_tree}.
The parameters used for the SVM method are listed
in Table \ref{SVM_parm}.  The parameter $\epsilon$ represents
the fitting tolerance for both AGF and SVM.
Parameters were hand-selected to maximize
efficiency without compromising accuracy using the cross-validation
procedure described above.

\label{comparison_results}

\begin{table}
  \begin{center}
  \caption{Summary of validation and comparison results}
     \begin{tabular}{|l|ccccc|} \hline
       Algorithm &     training &  classification & uncertainty &      accuracy &   correlation \\
                 &     time (s) &  time (s)       & coefficient &               & of R \\ \hline\hline
        Analytic &     N/A &  $1.29 \pm 0.20$ & $0.53 \pm 0.02$ & $0.906 \pm 0.005$ & $1.$ \\ 
             KNN &     N/A &  $5.59 \pm 0.31$ & $0.53 \pm 0.02$ & $0.905 \pm 0.005$ & $0.9953 \pm 0.0006$\\ 
             AGF &     N/A &  $9.40 \pm 0.31$ & $0.53 \pm 0.02$ & $0.905 \pm 0.005$ & $0.9979 \pm 0.0003$\\ 
     AGF borders & $4.19\pm0.10$ & $0.013\pm0.002$ & $0.53\pm0.02$ & $0.905\pm0.005$ & $0.9972 \pm 0.0008$\\ 
             LVQ & $2.81\pm0.01$ & $0.101\pm0.003$ & $0.50\pm0.02$ & $0.898\pm0.006$ &  N/A \\ 
             SVM & $112.4\pm3.6$ & $2.22\pm0.15$ & $0.53\pm0.02$ & $0.905\pm0.005$ & $0.9978\pm 0.0003$ \\ \hline
    \end{tabular}
    \label{comp_table}
  \end{center}
\end{table}

Table \ref{comp_table} shows the results of the comparison over twenty (20)
trials.  The uncertainty
coefficient is a better method of validating classification results
than simple accuracy (fraction of correct guesses) 
because it is not affected by the relative number of samples in each class.
It is defined as follows:
\begin{eqnarray}
  H(i|j) & = & - \sum_{i, j} P(i, j) \ln P(i|j)\\
  H(i) & = & \sum_i P(i) \ln P(i) \\
  U(i|j) & = & \frac{H(i)-H(i|j)}{H(i)}
\end{eqnarray}
where $i$ and $j$ enumerate the true and retrieved classes respectively, 
$P(i, j)$ is their joint probability, $P(i|j) = P(i, j)/P(j)$ is their
conditional probability, and $P(i) = \sum_j P(i, j)$ is the total,
invariant probability of the first class.  If we think of the classification
procedure as a noisy channel, then $U(i|j)$ quantifies how many bits of knowledge
we have of the true value of the class as a fraction of the maximum
number of bits it is possible to transmit per classification
\citep{nr_inc2,Shannon}.

\begin{figure}
\begin{center}
  \resizebox{0.9\textwidth}{!}{
    \includegraphics{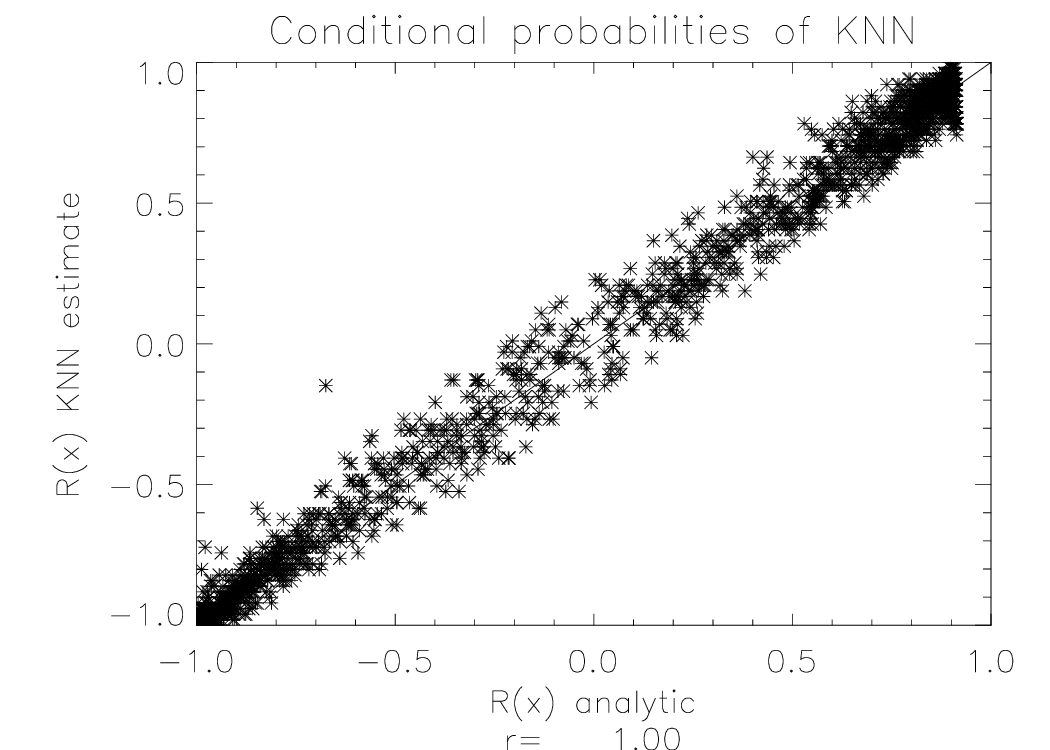}}
  \caption{Comparison between estimates of conditional probabilities
for the synthetic test classes.  $Y$-axis is KNN,
$X$-axis is semi-analytic.}
  \label{con_knn}
\end{center}
\end{figure}

\begin{figure}
\begin{center}
  \resizebox{0.9\textwidth}{!}{
    \includegraphics{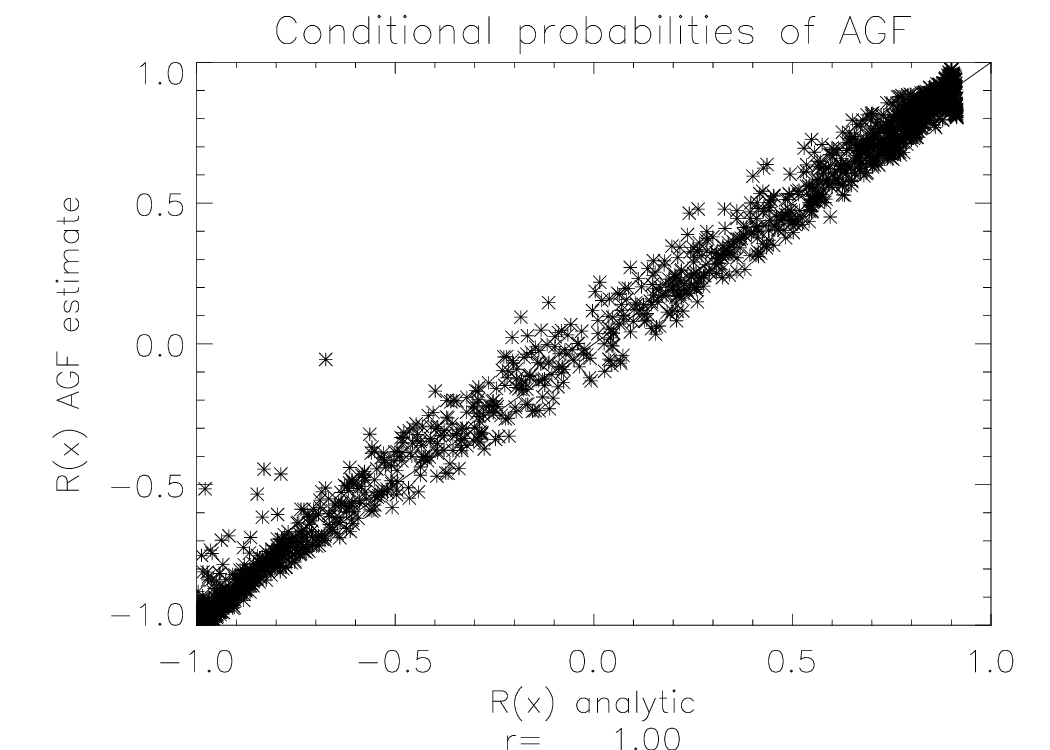}}
  \caption{Comparison between estimates of conditional probabilities
for the synthetic test classes.  $Y$-axis is AGF without borders training,
$X$-axis is semi-analytic.}
  \label{con_agf}
\end{center}
\end{figure}

\begin{figure}
\begin{center}
  \resizebox{0.9\textwidth}{!}{
    \includegraphics{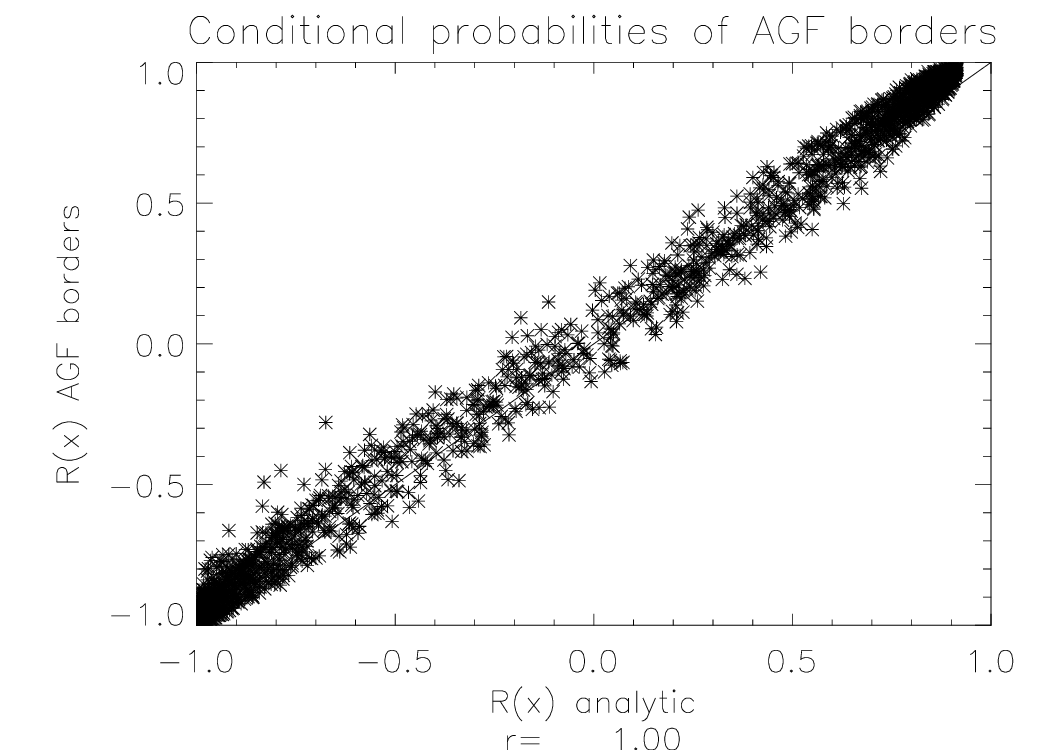}}
  \caption{Comparison between estimates of conditional probabilities
for the synthetic test classes.  $Y$-axis is AGF with borders training,
$X$-axis is semi-analytic.}
  \label{con_agf_bord}
\end{center}
\end{figure}

\begin{figure}
\begin{center}
  \resizebox{0.9\textwidth}{!}{
    \includegraphics{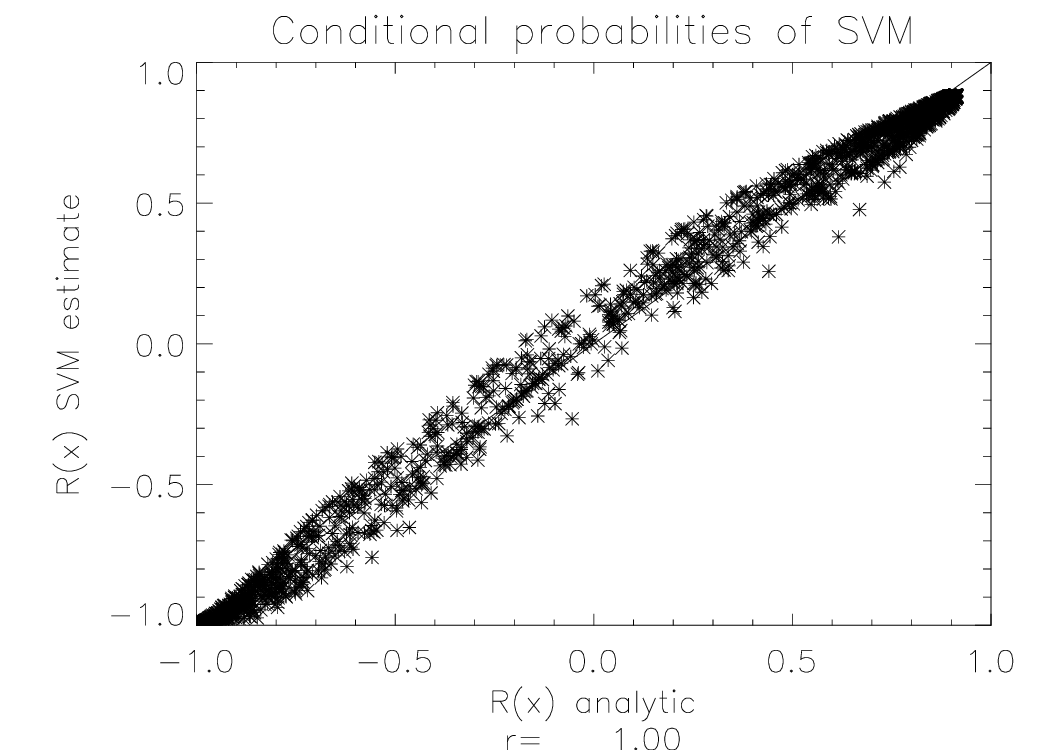}}
  \caption{Comparison between estimates of conditional probabilities
for the synthetic test classes.  $Y$-axis is LIBSVM,
$X$-axis is semi-analytic.}
  \label{con_svm}
\end{center}
\end{figure}

\begin{figure}
\begin{center}
  \resizebox{0.9\textwidth}{!}{
    \includegraphics{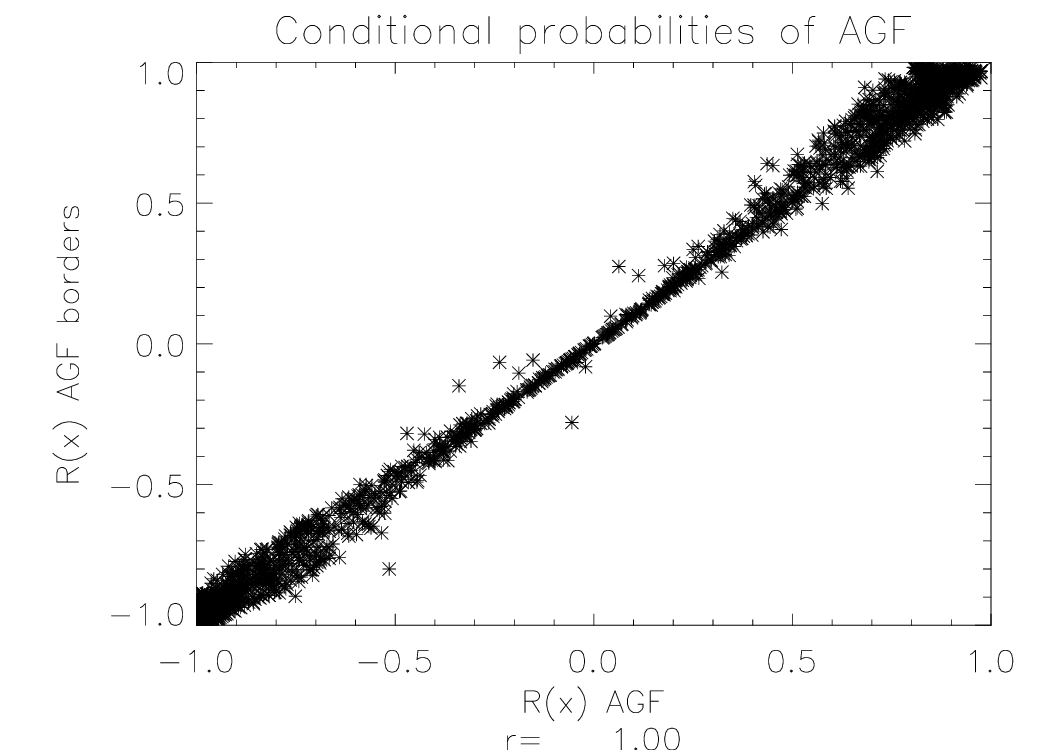}}
  \caption{Comparison between estimates of conditional probabilities
for the synthetic test classes.  $Y$-axis is AGF with borders training,
$X$-axis is AGF without.}
  \label{con_agf2}
\end{center}
\end{figure}

The last column in the table is simply the correlation coefficient of the
estimates of $R(\vecvar x)$ vs. the true values as computed by equations
(\ref{pdf_sc1a})--(\ref{pdf_sc2}).  For a visual comparison, see figures
\ref{con_knn}--\ref{con_svm}.  Figure \ref{con_agf2} compares
estimates from AGF with borders training and without.

The main thing to note from this comparison is how much faster
AGF with borders training is than the other four methods.
This is important if we need to process large amounts of 
satellite data, especially in real time.  The only other method
that even comes close is LVQ, especially in the training phase,
where it is much faster.
Unfortunately, it does not supply any knowledge of the
conditional probabilities, a necessary quantity for measuring
the accuracy of a given classification with no prior knowledge
of its true value.  They are also useful for re-calibration of retrieved
images, as will be demonstrated later.

\begin{figure}
\begin{center}
  \resizebox{0.95\textwidth}{!}{
    \includegraphics{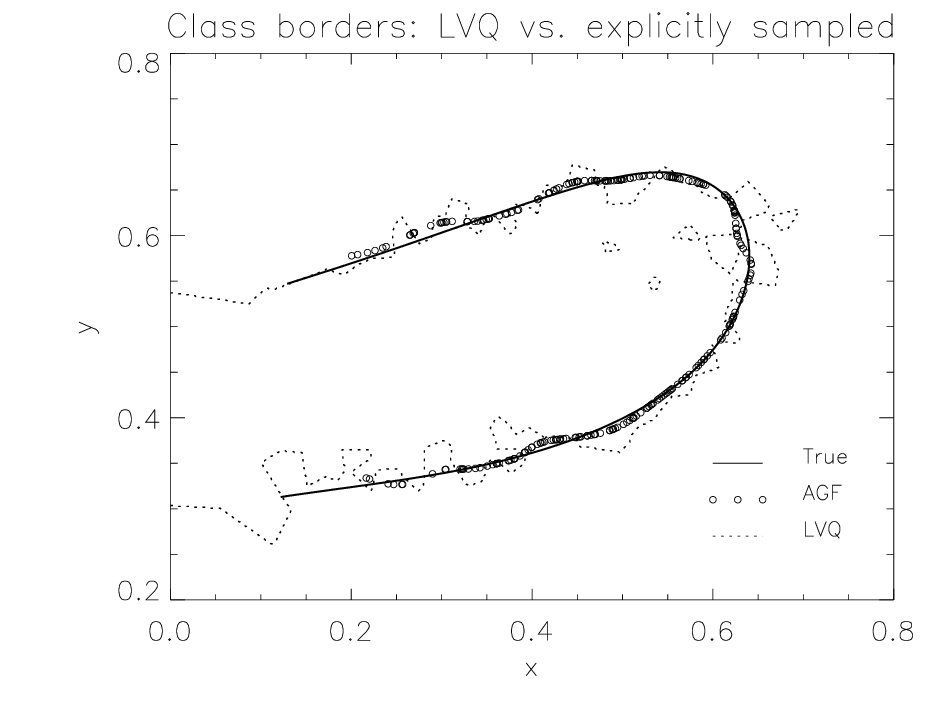}}
  \label{lvq_vs_brd}
  \caption{The class border of the synthetic test classes as characterized
by three different methods.}
\end{center}
\end{figure}

Another short-coming of the LVQ method is that it samples very
sparsely near the class borders, the exact region where we
require the most knowledge.  In fact the density of codebook
vectors approaches zero at the class border; the method is
actually designed this way!  When we plot the class border
between the codebook vectors from an LVQ training run, this
produces a line that is jagged and meandering.  Contrast this
to the border found via AGF as shown in figure \ref{lvq_vs_brd}.

\section{Application to Landsat images}

\begin{figure}
\begin{center}
\includegraphics[width=0.95\textwidth]{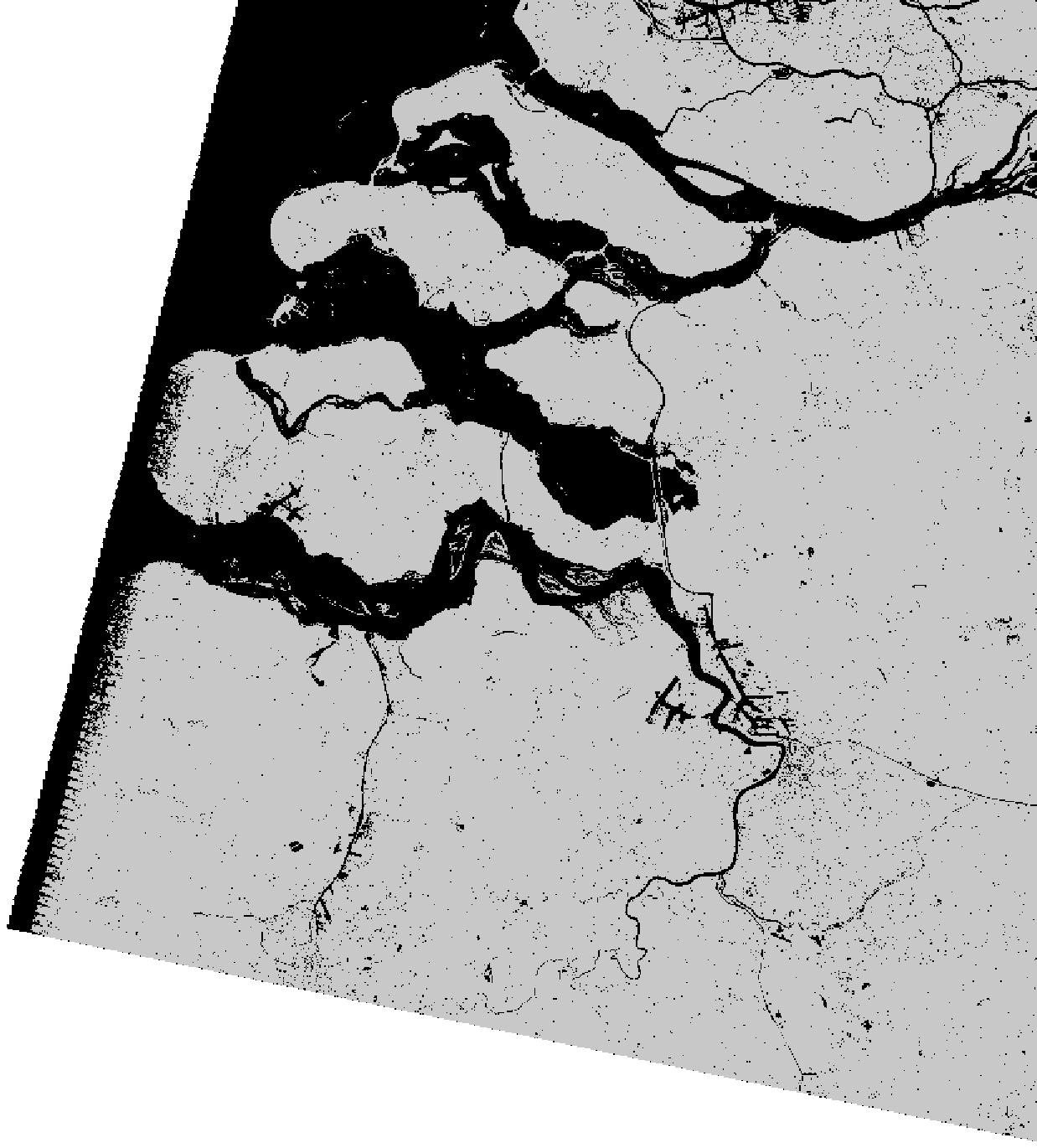}
\caption{Landsat image (LE71980242002267EDC00) of the S. Netherlands 
whose pixels have been automatically classified into land and water
using a training dataset derived from three other Landsat images.}
\label{LE71980242007217ASN00}
\end{center}
\end{figure}

Because of its global coverage and high resolution, 
Landsat 7 is one of the most sophisticated surface-mapping instruments.
It images the globe using seven channels in the visible and near-infrared.
Because of the 30 m resolution, any kind of processing will be
highly computationally intensive.  Landsat images are frequently used
``as-is'' to simulate aerial photographs 
by simply forming a colour image from three of the channels.
A more powerful use, however, is feature-detection or surface classification
using automated discrimination algorithms such as statistical classification.

The waterways in the Netherlands form a dense, complex network and 
because they produce dry land where once there was open sea are
one of the engineering wonders of the world.  Statistical classification
was used to map the lakes, canals and rivers of the Netherlands based
on a Landsat 7 image as seen in figure \ref{LE71980242007217ASN00}. 
Training data was selected manually from three images: two from
Southern Ontario -- LE70170292008280EDC00 and LE70180302008207EDC00 --
whose surface is one third fresh water and one from Northern Germany --
LE71960232008206ASN00 -- whose landscape resembles the Netherlands.
1822 training samples were selected in total while the six Landsat 7 channels with
30 m resolution were used in the analysis.  

\begin{figure}
\begin{center}
\includegraphics[width=0.9\textwidth]{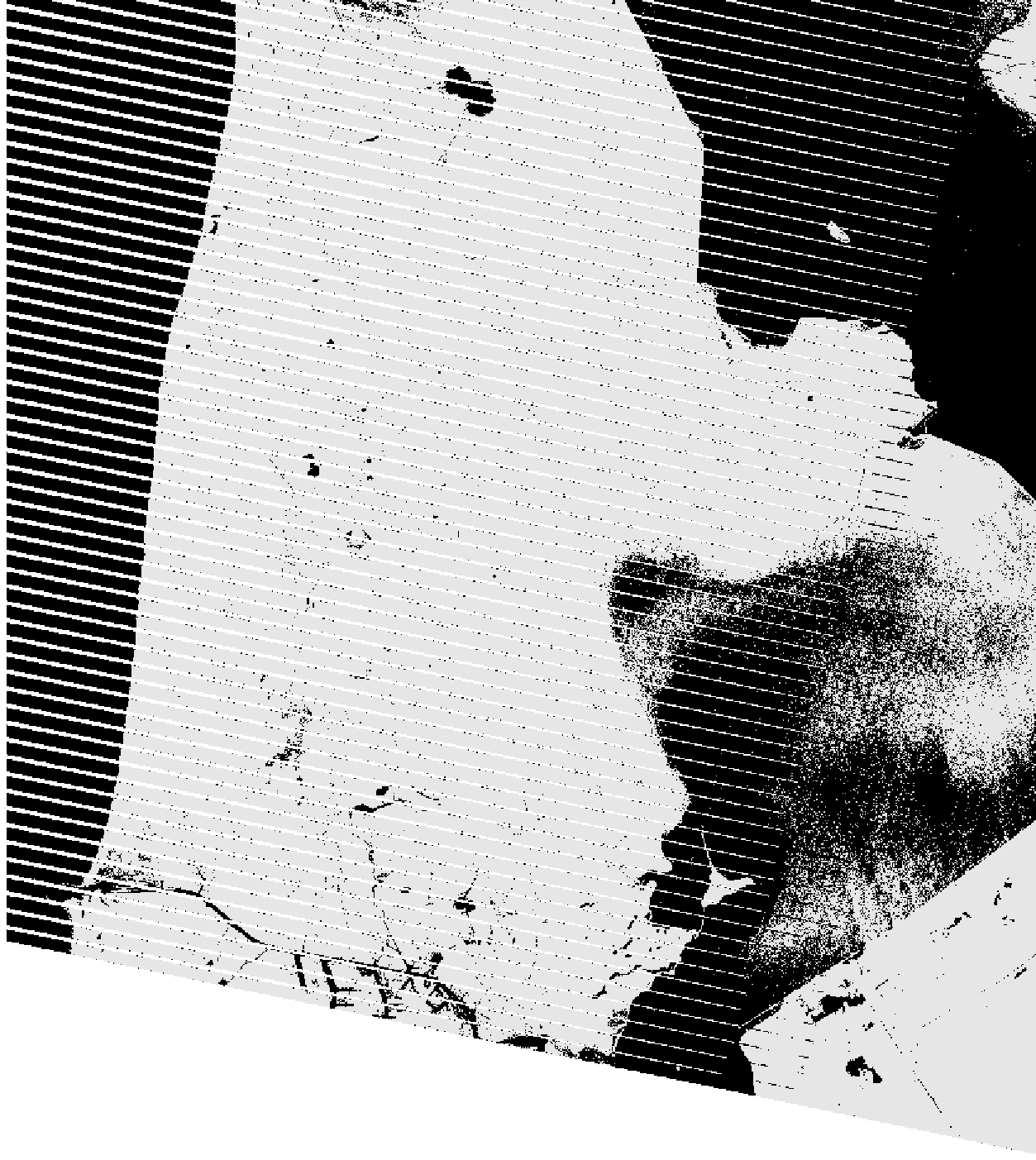}
\caption{Landsat image (LE71980232007217ASN00) of the N. Netherlands 
whose pixels have been automatically classified into land and water.
The horizontal striping is caused by a malfunction in the instrument
control system affecting all Landsat 7 images after 31 May 2003.}
\label{N_Neth_waterways}
\end{center}
\end{figure}

\begin{figure}
\begin{center}
\includegraphics[width=0.9\textwidth]{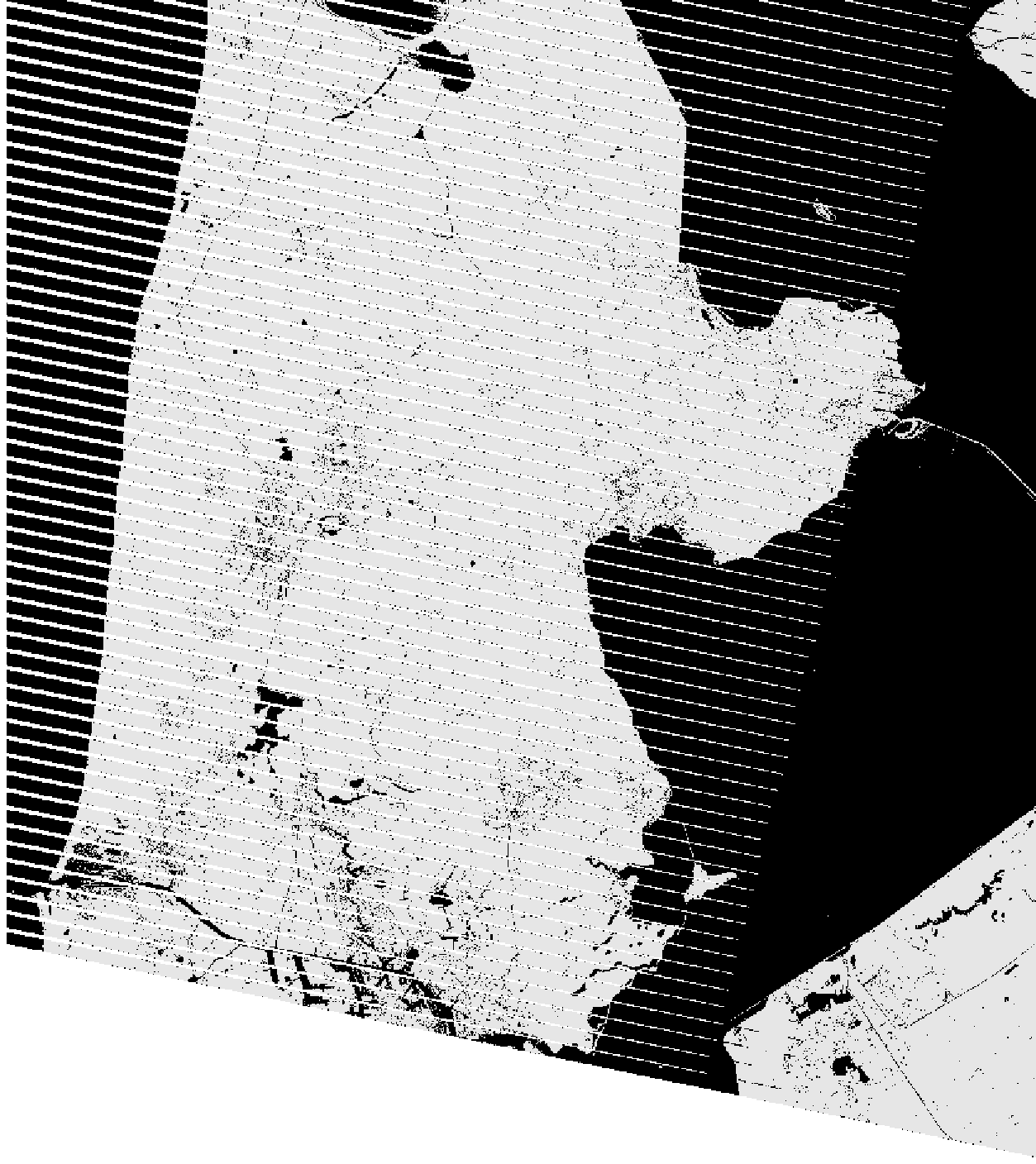}
\caption{Landsat image (LE71980232007217ASN00) of the N. Netherlands 
whose pixels have been automatically classified into land and water.
Image has been re-calibrated by shifting the discrimination border.}
\label{N_Neth_corrected}
\end{center}
\end{figure}

Training times for AGF, LVQ, SVM and SVM with probability estimates were
0.5 s, 0.7 s, 0.15 s and 0.7 s, respectively.
Classification times were 5.5 minutes, 20 minutes, 55
minutes and 58 minutes, respectively.
In this example the training time is fairly immaterial because of the
small number of samples, however for classification, AGF is nevertheless
still the clear winner.  200 border samples were used.  For LVQ,
200 ``codebook'' vectors were employed, 
although as implied in section \ref{comparison_results}, 
this is giving LVQ an
unfair speed advantage since it will take more ``codebook'' vectors than
border samples to represent the discrimination border to a similar 
level of precision.

\begin{figure}
\begin{center}
\includegraphics[width=0.9\textwidth]{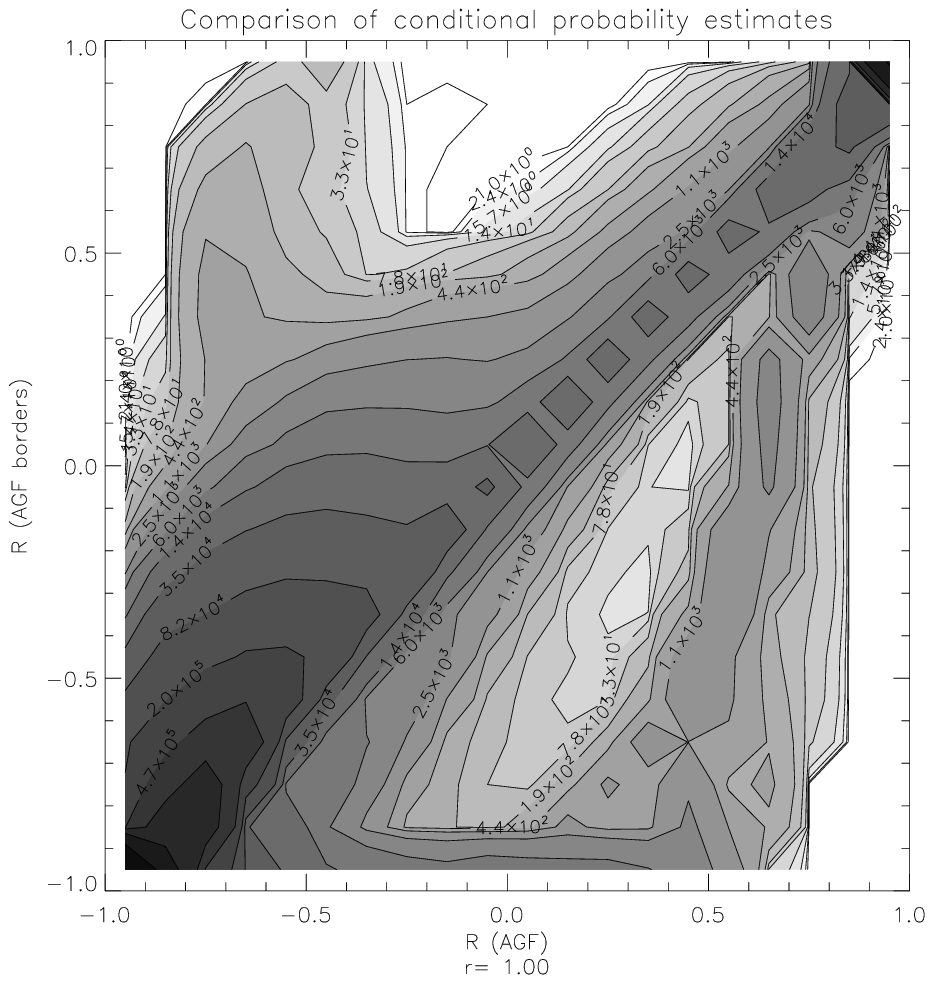}
\caption{Comparison of conditional probability estimates using AGF with
borders training versus direct AGF of pixels classified into land and water
from Landsat 7 image LE71980242002267EDC00.
Contours follow a geometric progression.}
\label{con_comp_ls7}
\end{center}
\end{figure}

Figure \ref{N_Neth_waterways} demonstrates the utility of having the
conditional probabilities available.  
Using a different set of lower-quality training data, 
many water points are now mis-classified as land.  
Therefore, we re-calibrate
the algorithm by choosing a different, lower threshold value, $R=-0.8$, for
the discrimination border, similar to what is described in \citet{Mills2009}.  
The corrected image is shown in figure \ref{N_Neth_corrected}.
It is still not perfect, but the transformation has done a good job
of re-classifying many of the points lying in the tidal flats as water
instead of land.

In real-world problems, if the PDF's of the classes are not monotonic
and roughly Gaussian near the borders, then conditional probability
estimates based on (\ref{confidence_est}) will be inaccurate.
Fortunately, this is rarely the case.  Figure \ref{con_comp_ls7} 
compares conditional probabilities estimated
by extrapolation using equation (\ref{confidence_est})
and more directly using the right-most side of equation (\ref{Rdef})
for the surface classifications illustrated in figure \ref{LE71980242007217ASN00}.
In libSVM, conditional probabilities are also estimated using a one-dimensional
parametrisation \citep{CC01a}, so the best way to ensure accurate estimates
is to apply one of the more direct methods -- KNN or AGF without borders
training.

\section{Discussion}

\subsection{Algorithmic efficiency}

The speed of the competing algorithms for a single problem does not
tell the whole story since we also need to know the algorithmic
efficiency, that is how the speed of the different algorithms
depends on both the training set size and on the selection of parameters.
In the case of KNN,
classification times depend on the time it takes to select the $k$
neighbours, in our case $n \log k$ time.  Since $k$ is generally increased
with increasing numbers of training samples, this means that the
time dependence is more like $n \log n$.  Actual tests of the selection
algorithm, however, show a very weak dependence on $k$.
Best case performance for a selection algorithm such as ours 
that sorts the results is actually $n + k \log k$, 
while for one that does not it is $n$ \citep{Knuth1998}.

If this step is included, 
the bottleneck in the AGF algorithm is selecting the $k$ nearest neighbours,
hence the time efficiency will also be roughly $n \log k$ worst case,
although with further overhead of order $k$ from calculating the weights.
If the selection step is skipped, the time efficiency will be $n$, but
with a large coefficient.  Time efficiency for LVQ training will be $n$,
while for SVM training, it is $n^2$ since the solution of
a matrix equation is required.  Note that all methods will have extra
overhead from reading in the training data and also from writing
the output.

For algorithms that have a separate training phase, there is also the
issue of the classification time.  For both LVQ and AGF borders,
classification times will be independent of the number of training samples.
Time efficiency will rather depend linearly on the number of samples
used to represent the discrimination border
-- codebook vectors for LVQ and border samples for AGF -- 
which are both adjustable parameters.  
The larger these parameters, the greater
the accuracy, although with diminishing returns.  
For equivalent accuracy, 
more LVQ codebook vectors than AGF border samples will be needed.

In the case of SVM, classification times \textit{do} depend on the
number of training samples, which would seem to make the training
phase a bit redundant.  The time efficiency appears to be a bit
better than $n$, where $n$ is the number of training samples.

\subsection{Use for continuum retrieval}

\begin{figure}
\begin{center}
\includegraphics[angle=90,width=0.9\textwidth]{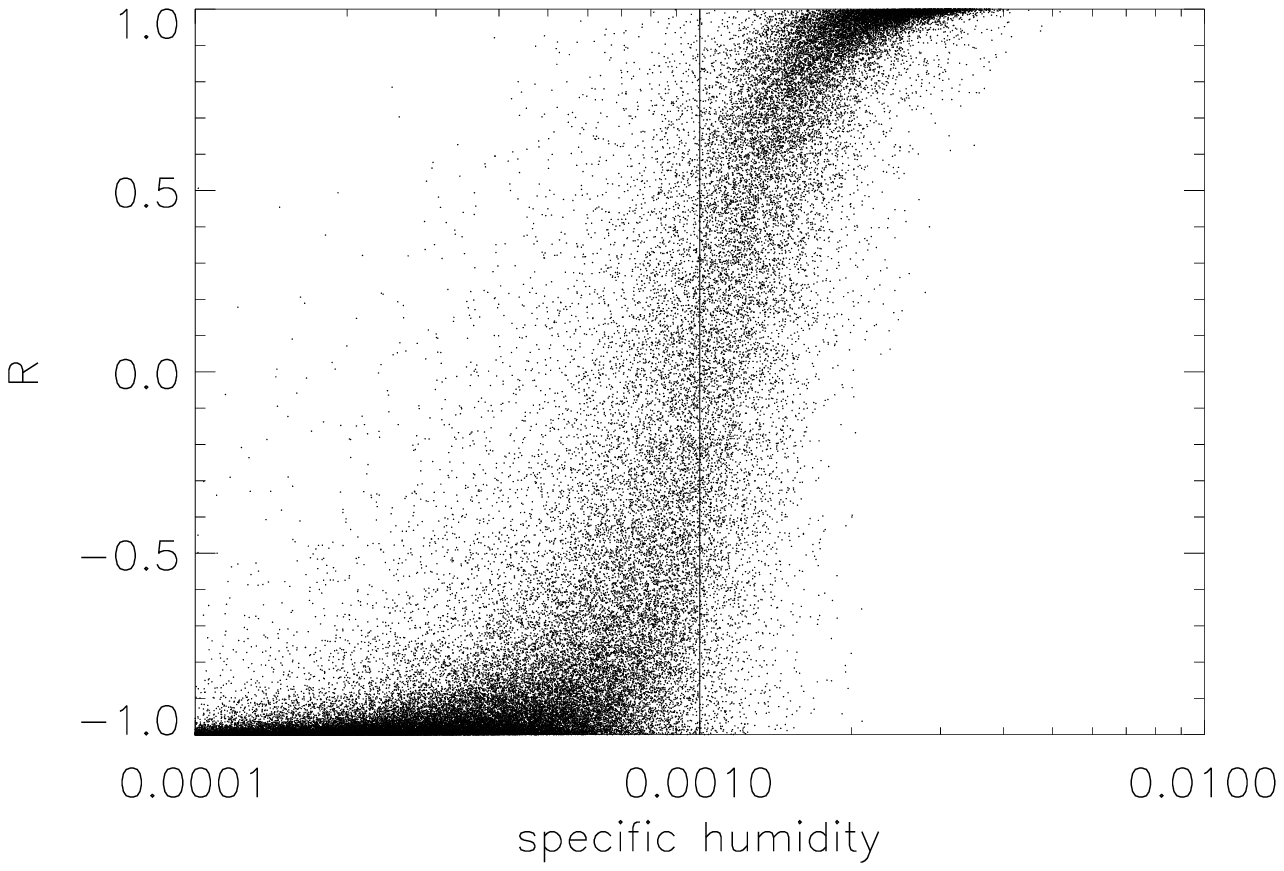}
\caption{Comparison of conditional probability estimates with continuum
values for a discrete water-vapour retrieval (threshold value for two ranges is set at 
0.001 mass-mixing ratio).
Conditional probabilities are estimated directly using AGF without borders training.}
\label{qvsr}
\end{center}
\end{figure}

Within the context of satellite remote sensing, statistical classification
would appear to be a somewhat specialized tool, useful chiefly for processing
images by extracting features or classifying surface types.  The method
truly comes into its own, however, when recognised as an
efficient, general, non-linear inverse method.

Like a neural network, statistical classification generates a direct
inverse method that has the potential to be very fast.  Unlike a neural
network, however, there is no possibility of getting stuck in a local
minimum when trying optimise the model.  Accuracy of the discrimination
border is limited only by the resolution of the samples.  The presence of
local minima may also confound inverse methods,
such as optimal estimation \citep{Rodgers2000}, 
that use a forward model directly.

Also unlike a neural network, statistical classification generates only
discrete values: a limitation which is surprisingly easy to overcome.
A continuum variable can be retrieved by dividing it into
ranges.  Designate the continuum variable as $q$.
Just like in the discrete case, the relationship between the measurement variable,
$\vecvar x$, and the state variable, $q$, will be governed by a conditional
probability, $P(q | \vecvar x)$.  Suppose we divide $q$ into two ranges
with a threshold value, $q_0$, so that the classes are defined as follows:
\begin{equation}
c = \left \lbrace \begin{array}{lc} 1; & q < q_0 \\ 
                2; & q \geq q_0\end{array} \right .
\end{equation}
The continuum conditional probability is transformed
to the discrete conditional probability by integration:
\begin{eqnarray}
P(1 | \vecvar{x}) = \int_{-\infty}^{q_0} P(q | \vecvar{x}) dq
\label{pq2pc1}\\
P(2 | \vecvar{x}) =  \int^{\infty}_{q_0} P(q | \vecvar{x}) dq
\label{pq2pc2}
\end{eqnarray}

If the divisions in $q$ are made fine enough, i.e. smaller than the estimated error,
then these can be used directly.  Alternatively, the continuum value can
be reconstituted in a number of other ways.  For a two-dimensional retrieval,
we can retrieve a series of isolines and then interpolate between them
\citep{Mills2009}.  If conditional probabilities are available, they can actually
make quite a good proxy for the continuum result.  Figure \ref{qvsr}
compares conditional probabilities with continuum values 
from the water-vapour retrieval described in \citet{Mills2009}.
It is easy to show from equations (\ref{pq2pc1}) and (\ref{pq2pc2})
that if the statistics of $q$ for a given $\vecvar x$
in $P(q | \vecvar x)$ are Gaussian, then the relationship
between the two will be an error function, as seen in the figure.
Given:
\begin{equation}
P(q | \vecvar x) = \frac{1}{\sqrt{2 \pi} \sigma_q} 
	\exp \left \lbrace - \frac{\left [\bar q (\vecvar x) - q\right ]^2}{2 \sigma_q} \right \rbrace
\end{equation}
where $\sigma_q$ is the width of the distribution and $\bar q$ is the expectation value.
Then:
\begin{equation}
R = \mathrm{erf} \left [ \frac{\bar q (\vecvar x) - q_0}{\sqrt 2 \sigma_q} \right ]
\end{equation}
This result can be used not only to estimate the continuum result,
but also to replace the hyperbolic tangent form used in equation (\ref{confidence_est})
to estimate $R$.  This assumes that $\bar q$ is roughly linear in $\vecvar x$
near the class border.

Another feature of such a retrieval is its robustness.  Once classes have been
defined, it is impossible to generate a result outside of this range.
Moreover, it is easy to see from (\ref{pq2pc1}) and (\ref{pq2pc2}) 
that classification retrieval of continuum values
is what is known as a robust estimator.  Normally, $q$ is estimated by evaluating its
expectation value or first moment:
\begin{equation}
\bar q = \int_{-\infty}^{\infty} P(q | \vecvar x) q dq
\end{equation}
By contrast, the location of $q_0$ is found by setting: 
\begin{equation}
\int_{-\infty}^{q_0} P(q | \vecvar{x}) dq = 
\int^{\infty}_{q_0} P(q | \vecvar{x}) dq
\end{equation}
or equalizing the fraction of 
``zeroth order'' moment on each side of the threshold.  
This type of formulation is characteristic of robust estimators 
and results in outliers in the training data 
having less effect on the final model \citep{nr_inc2}.

Finally, with the continuum variable divided into multiple ranges, 
returned error statistics can be made more detailed.  By examining the
conditional probabilities within each range, not only the statistics, but 
also the approximate functional form of the error may be determined.

\section{Conclusion}

A statistical classification algorithm based on kernel-density 
estimation was described, which we term Adaptive Gaussian Filtering or AGF.  
The many refinements of this algorithm produce
impressive speed gains compared to other methods making it appropriate for
processing large amounts of satellite data, especially in real-time.
As applied to a synthetic test dataset, 
the method is shown to be 25 times as fast for training 
and 125 times as fast for classification, when compared to LIBSVM, a Support
Vector Machine implementation.  The performance advantage becomes greater the
larger the number of training samples.  Therefore, when tested on real data,
discriminating land from water in the Netherlands, the performance advantage
was reduced because of the smaller training dataset 
 -- training times were roughly equivalent, while classification
was ten times as fast.

The algorithm was also compared to $k$-nearest-neighbours (KNN) and Kohonen's Learning
Vector Quantisation (LVQ).  Three steps can be distinguished in the
AGF algorithm.  First, the conditional probabilities are estimated using kernel
averaging.  These estimates can be used directly to classify a set of test
data from the training data, in a manner equivalent to a KNN.  
Second, the conditional probability estimates are used to search for the 
discrimination or class border in the case of a two-class classification.
Once the discrimination border is known, it can be used to generate classification
estimates more quickly than directly with the training data.
Thus, KNN is not as fast as AGF when borders training is included.

While LVQ is similarly fast, both for training and classification, 
it does not return estimates of the conditional
probabilities and is less accurate.  Estimates of the conditional probabilities
are useful for determining the accuracy of an estimate in absence of
its true value.  They are also useful for recalibrating an image when the
classification estimates are biased.

Conditional probabilities are estimated using a simple parametrisation based on 
their gradients at the border.
For the most accurate estimates of the conditional probabilities, a direct
method such as KNN or AGF without borders training can be used.
For most applications, however, the gains in accuracy will be 
too small to offset the significant speed penalty.

While statistical classification finds broad application in image processing
techniques like feature extraction and surface detection, its full power as a
general, non-linear inverse method has yet to be harnessed.  Continuum variables
can be retrieved by dividing them into discrete ranges.  Such a technique has
the advantage over a neural network or a more direct inverse method such as
optimal estimation in that there is no possibility of becoming stuck in a local
minimum.  The accuracy of the estimates is limited only by the number and
resolution of the training samples.

Software can be found at: \url{http://libagf.sourceforge.net}.

\section*{Acknowledgements}

Thank you very much to my colleagues from the IUP, University of
Bremen for their continued support and encouragement of this work,
in particular Stefan Buehler, Georg Heygster and Oliver Lemke.  
Thanks to Christian Melsheimer for his very valued comments on the draft manuscript.

This work was partially funded by the German Federal Ministry of Education and
Research (BMBF), within the AFO2000 project UTH-MOS, grant 07ATC04. It
is a contribution to COST Action 723 `Data Exploitation and Modeling
for the Upper Troposphere and Lower Stratosphere'.


\bibliography{agf2}


\appendix

\section{Gradient of an AGF estimate}
\footnote{This section does not appear in the original article.}

\label{appendix}

Since the final result is just a summation of the filter weights, 
 multiplied by some nominally constant values, we start by taking the gradients of the weights: 

\begin{equation}
  \frac{\partial w_i}{\partial x_j} = \left (\frac{\partial w_i}{\partial x_j} \right )_\sigma
		+ \frac{\partial w_i}{\partial \sigma} \frac{\partial \sigma}{\partial x_j}
  \label{gradagf:grad_w}
\end{equation}

Since the filter width is not constant, we must include a term to account for its change.
The second factor of the second term, the gradient of the filter width, is not known,
but we can easily solve for it by taking the derivative of the total weight, which is
a constant:

\begin{eqnarray}
  \frac{\partial W}{\partial x_j} & = & \frac{\partial}{\partial x_j} \sum_i w_i \\
  & = & \sum_i \frac{\partial w_i}{\partial x_j} \\
  & = & \sum_i \left [\left (\frac{\partial w_i}{\partial x_j} \right )_\sigma
                + \frac{\partial w_i}{\partial \sigma} 
		\frac{\partial \sigma}{\partial x_j} \right ] \\
  & = & 0
\end{eqnarray}
Solving:

\begin{equation}
  \frac{\partial \sigma}{\partial x_j} = - \frac
		{\sum_i \left (\frac{\partial w_i}{\partial x_j} \right )_\sigma}
		{\sum_i \frac{\partial w_i}{\partial \sigma}}
  \label{gradagf:grad_sigma}
\end{equation}

For the case of a Gaussian filter the weights are given by:

\begin{eqnarray}
  w_i & = & \exp \left [ - \frac{d_i^2}{2 \sigma^2} \right ] \\
  	& = & \exp \left [ - \frac{\sum_j (x_{ij} - x_j)^2}{2 \sigma^2} \right ]
\end{eqnarray}
which generates the following partials:
\begin{eqnarray}
  \left (\frac{\partial w_i}{\partial x_j} \right )_\sigma & = &
		 \frac{(x_{ij} - x_j)}{\sigma^2} w_i \\
  {\frac{\partial w_i}{\partial \sigma}} & = & \frac{d_i^2}{\sigma^3} w_i
\end{eqnarray}

Subsituting these into equations (\ref{gradagf:grad_sigma}) and 
(\ref{gradagf:grad_w}) respectively, produces the following:

\begin{eqnarray}
  \frac{\partial \sigma}{\partial x_j} & = & - \frac{\sigma \sum_i (x_{ij} - x_j) w_i}
		{\sum_i d_i^2 w_i} \\
  \frac{\partial w_i}{\partial x_j} & = & \frac{w_i}{\sigma^2} \left [x_{ij} - x_j 
		- d_i^2 \frac{\sum_k (x_{kj} - x_j) w_k} {\sum_k d_k^2 w_k} \right ]
\end{eqnarray}

For a two class classification, the gradient of the difference of the conditional 
probabilities is given as follows:

\begin{eqnarray}
  \frac{\partial R}{\partial x_j} & \approx & \frac{1}{W} \sum_i \frac{\partial w_i}{\partial x_j} (2c_i-3) \\
  & \approx & \frac{1}{W \sigma^2} \sum_i w_i (2c_i - 3) \left [ x_{ij} - x_j
                - d_i^2 \frac{\sum_k (x_{kj} - x_j) w_k} {\sum_k d_k^2 w_k} \right ]
\end{eqnarray}
using our rather awkward convention of enumerating the first class by ``$1$'' 
and the second by ``$2$.''

\end{document}